\documentclass[aps,prd,preprint,superscriptaddress,amsfonts,amssymb,amsmath,natbib,nofootinbib]{revtex4}
\pdfoutput=1
\usepackage{color}
\usepackage{times}
\usepackage{graphicx}
\usepackage{fancyhdr}
\usepackage{float}
\usepackage{subfigure}
\usepackage{acronym}
\usepackage{multirow}
\usepackage{array}

\usepackage{natbib}

\usepackage{amsmath}
\usepackage{amsfonts}
\usepackage{amssymb}
\usepackage{epsfig}
\usepackage[english]{babel}
\usepackage{rotating}
\usepackage{float}
\usepackage{graphics}

\begin{document}

\title{Upper limits on a stochastic gravitational-wave background
using LIGO and Virgo interferometers at 600-1000 Hz
}





\affiliation{LIGO - California Institute of Technology, Pasadena, CA  91125, USA$^\ast$}
\affiliation{California State University Fullerton, Fullerton CA 92831 USA$^\ast$}
\affiliation{SUPA, University of Glasgow, Glasgow, G12 8QQ, United Kingdom$^\ast$}
\affiliation{Laboratoire d'Annecy-le-Vieux de Physique des Particules (LAPP), Universit\'e de Savoie, CNRS/IN2P3, F-74941 Annecy-Le-Vieux, France$^\dagger$}
\affiliation{INFN, Sezione di Napoli $^a$; Universit\`a di Napoli 'Federico II'$^b$ Complesso Universitario di Monte S.Angelo, I-80126 Napoli; Universit\`a di Salerno, Fisciano, I-84084 Salerno$^c$, Italy$^\dagger$}
\affiliation{LIGO - Livingston Observatory, Livingston, LA  70754, USA$^\ast$}
\affiliation{Albert-Einstein-Institut, Max-Planck-Institut f\"ur Gravitationsphysik, D-30167 Hannover, Germany$^\ast$}
\affiliation{Leibniz Universit\"at Hannover, D-30167 Hannover, Germany$^\ast$}
\affiliation{Nikhef, Science Park, Amsterdam, the Netherlands$^a$; VU University Amsterdam, De Boelelaan 1081, 1081 HV Amsterdam, the Netherlands$^b$$^\dagger$}
\affiliation{National Astronomical Observatory of Japan, Tokyo  181-8588, Japan$^\ast$}
\affiliation{University of Wisconsin--Milwaukee, Milwaukee, WI  53201, USA$^\ast$}
\affiliation{University of Florida, Gainesville, FL  32611, USA$^\ast$}
\affiliation{University of Birmingham, Birmingham, B15 2TT, United Kingdom$^\ast$}
\affiliation{INFN, Sezione di Roma$^a$; Universit\`a 'La Sapienza'$^b$, I-00185 Roma, Italy$^\dagger$}
\affiliation{LIGO - Hanford Observatory, Richland, WA  99352, USA$^\ast$}
\affiliation{Albert-Einstein-Institut, Max-Planck-Institut f\"ur Gravitationsphysik, D-14476 Golm, Germany$^\ast$}
\affiliation{Montana State University, Bozeman, MT 59717, USA$^\ast$}
\affiliation{European Gravitational Observatory (EGO), I-56021 Cascina (PI), Italy$^\dagger$}
\affiliation{Syracuse University, Syracuse, NY  13244, USA$^\ast$}
\affiliation{LIGO - Massachusetts Institute of Technology, Cambridge, MA 02139, USA$^\ast$}
\affiliation{Laboratoire AstroParticule et Cosmologie (APC) Universit\'e Paris Diderot, CNRS: IN2P3, CEA: DSM/IRFU, Observatoire de Paris, 10 rue A.Domon et L.Duquet, 75013 Paris - France$^\dagger$}
\affiliation{Columbia University, New York, NY  10027, USA$^\ast$}
\affiliation{INFN, Sezione di Pisa$^a$; Universit\`a di Pisa$^b$; I-56127 Pisa; Universit\`a di Siena, I-53100 Siena$^c$, Italy$^\dagger$}
\affiliation{Stanford University, Stanford, CA  94305, USA$^\ast$}
\affiliation{IM-PAN 00-956 Warsaw$^a$; Astronomical Observatory Warsaw University 00-478 Warsaw$^b$; CAMK-PAN 00-716 Warsaw$^c$; Bia{\l}ystok University 15-424 Bia{\l}ystok$^d$; IPJ 05-400 \'Swierk-Otwock$^e$; Institute of Astronomy 65-265 Zielona G\'ora$^f$,  Poland$^\dagger$}
\affiliation{The University of Texas at Brownsville and Texas Southmost College, Brownsville, TX  78520, USA$^\ast$}
\affiliation{San Jose State University, San Jose, CA 95192, USA$^\ast$}
\affiliation{Moscow State University, Moscow, 119992, Russia$^\ast$}
\affiliation{LAL, Universit\'e Paris-Sud, IN2P3/CNRS, F-91898 Orsay$^a$; ESPCI, CNRS,  F-75005 Paris$^b$, France$^\dagger$}
\affiliation{NASA/Goddard Space Flight Center, Greenbelt, MD  20771, USA$^\ast$}
\affiliation{University of Western Australia, Crawley, WA 6009, Australia$^\ast$}
\affiliation{The Pennsylvania State University, University Park, PA  16802, USA$^\ast$}
\affiliation{Universit\'e Nice-Sophia-Antipolis, CNRS, Observatoire de la C\^ote d'Azur, F-06304 Nice$^a$; Institut de Physique de Rennes, CNRS, Universit\'e de Rennes 1, 35042 Rennes$^b$, France$^\dagger$}
\affiliation{Laboratoire des Mat\'eriaux Avanc\'es (LMA), IN2P3/CNRS, F-69622 Villeurbanne, Lyon, France$^\dagger$}
\affiliation{Washington State University, Pullman, WA 99164, USA$^\ast$}
\affiliation{INFN, Sezione di Perugia$^a$; Universit\`a di Perugia$^b$, I-06123 Perugia,Italy$^\dagger$}
\affiliation{INFN, Sezione di Firenze, I-50019 Sesto Fiorentino$^a$; Universit\`a degli Studi di Urbino 'Carlo Bo', I-61029 Urbino$^b$, Italy$^\dagger$}
\affiliation{University of Oregon, Eugene, OR  97403, USA$^\ast$}
\affiliation{Laboratoire Kastler Brossel, ENS, CNRS, UPMC, Universit\'e Pierre et Marie Curie, 4 Place Jussieu, F-75005 Paris, France$^\dagger$}
\affiliation{University of Maryland, College Park, MD 20742 USA$^\ast$}
\affiliation{University of Massachusetts - Amherst, Amherst, MA 01003, USA$^\ast$}
\affiliation{Canadian Institute for Theoretical Astrophysics, University of Toronto, Toronto, Ontario, M5S 3H8, Canada$^\ast$}
\affiliation{Tsinghua University, Beijing 100084 China$^\ast$}
\affiliation{University of Michigan, Ann Arbor, MI  48109, USA$^\ast$}
\affiliation{Louisiana State University, Baton Rouge, LA  70803, USA$^\ast$}
\affiliation{The University of Mississippi, University, MS 38677, USA$^\ast$}
\affiliation{Charles Sturt University, Wagga Wagga, NSW 2678, Australia$^\ast$}
\affiliation{Caltech-CaRT, Pasadena, CA  91125, USA$^\ast$}
\affiliation{INFN, Sezione di Genova;  I-16146  Genova, Italy$^\dagger$}
\affiliation{Pusan National University, Busan 609-735, Korea$^\ast$}
\affiliation{Australian National University, Canberra, ACT 0200, Australia$^\ast$}
\affiliation{Carleton College, Northfield, MN  55057, USA$^\ast$}
\affiliation{The University of Melbourne, Parkville, VIC 3010, Australia$^\ast$}
\affiliation{Cardiff University, Cardiff, CF24 3AA, United Kingdom$^\ast$}
\affiliation{INFN, Sezione di Roma Tor Vergata$^a$; Universit\`a di Roma Tor Vergata, I-00133 Roma$^b$; Universit\`a dell'Aquila, I-67100 L'Aquila$^c$, Italy$^\dagger$}
\affiliation{University of Salerno, I-84084 Fisciano (Salerno), Italy and INFN (Sezione di Napoli), Italy$^\ast$}
\affiliation{The University of Sheffield, Sheffield S10 2TN, United Kingdom$^\ast$}
\affiliation{RMKI, H-1121 Budapest, Konkoly Thege Mikl\'os \'ut 29-33, Hungary$^\dagger$}
\affiliation{INFN, Gruppo Collegato di Trento$^a$ and Universit\`a di Trento$^b$,  I-38050 Povo, Trento, Italy;   INFN, Sezione di Padova$^c$ and Universit\`a di Padova$^d$, I-35131 Padova, Italy$^\dagger$}
\affiliation{Inter-University Centre for Astronomy and Astrophysics, Pune - 411007, India$^\ast$}
\affiliation{California Institute of Technology, Pasadena, CA  91125, USA$^\ast$}
\affiliation{Northwestern University, Evanston, IL  60208, USA$^\ast$}
\affiliation{University of Cambridge, Cambridge, CB2 1TN, United Kingdom$^\ast$}
\affiliation{The University of Texas at Austin, Austin, TX 78712, USA$^\ast$}
\affiliation{E\"otv\"os Lor\'and University, Budapest, 1117 Hungary$^\ast$}
\affiliation{University of Szeged, 6720 Szeged, D\'om t\'er 9, Hungary$^\ast$}
\affiliation{Universitat de les Illes Balears, E-07122 Palma de Mallorca, Spain$^\ast$}
\affiliation{Rutherford Appleton Laboratory, HSIC, Chilton, Didcot, Oxon OX11 0QX United Kingdom$^\ast$}
\affiliation{Embry-Riddle Aeronautical University, Prescott, AZ   86301 USA$^\ast$}
\affiliation{National Institute for Mathematical Sciences, Daejeon 305-390, Korea$^\ast$}
\affiliation{Perimeter Institute for Theoretical Physics, Ontario, N2L 2Y5, Canada$^\ast$}
\affiliation{University of New Hampshire, Durham, NH 03824, USA$^\ast$}
\affiliation{University of Adelaide, Adelaide, SA 5005, Australia$^\ast$}
\affiliation{University of Southampton, Southampton, SO17 1BJ, United Kingdom$^\ast$}
\affiliation{University of Minnesota, Minneapolis, MN 55455, USA$^\ast$}
\affiliation{Korea Institute of Science and Technology Information, Daejeon 305-806, Korea$^\ast$}
\affiliation{Hobart and William Smith Colleges, Geneva, NY  14456, USA$^\ast$}
\affiliation{Institute of Applied Physics, Nizhny Novgorod, 603950, Russia$^\ast$}
\affiliation{Lund Observatory, Box 43, SE-221 00, Lund, Sweden$^\ast$}
\affiliation{Hanyang University, Seoul 133-791, Korea$^\ast$}
\affiliation{Seoul National University, Seoul 151-742, Korea$^\ast$}
\affiliation{University of Strathclyde, Glasgow, G1 1XQ, United Kingdom$^\ast$}
\affiliation{Southern University and A\&M College, Baton Rouge, LA  70813, USA$^\ast$}
\affiliation{University of Rochester, Rochester, NY  14627, USA$^\ast$}
\affiliation{Rochester Institute of Technology, Rochester, NY  14623, USA$^\ast$}
\affiliation{University of Sannio at Benevento, I-82100 Benevento, Italy and INFN (Sezione di Napoli), Italy$^\ast$}
\affiliation{Louisiana Tech University, Ruston, LA  71272, USA$^\ast$}
\affiliation{McNeese State University, Lake Charles, LA 70609 USA$^\ast$}
\affiliation{Andrews University, Berrien Springs, MI 49104 USA$^\ast$}
\affiliation{Trinity University, San Antonio, TX  78212, USA$^\ast$}
\affiliation{Southeastern Louisiana University, Hammond, LA  70402, USA$^\ast$}
\author{J.~Abadie$^\text{1}$}\noaffiliation\author{B.~P.~Abbott$^\text{1}$}\noaffiliation\author{R.~Abbott$^\text{1}$}\noaffiliation\author{T.~D.~Abbott$^\text{2}$}\noaffiliation\author{M.~Abernathy$^\text{3}$}\noaffiliation\author{T.~Accadia$^\text{4}$}\noaffiliation\author{F.~Acernese$^\text{5a,5c}$}\noaffiliation\author{C.~Adams$^\text{6}$}\noaffiliation\author{R.~Adhikari$^\text{1}$}\noaffiliation\author{C.~Affeldt$^\text{7,8}$}\noaffiliation
% ...

\collaboration{$^\ast$The LIGO Scientific Collaboration and $^\dagger$The Virgo Collaboration}
\noaffiliation





\begin{abstract}\quad
A stochastic background of gravitational waves is expected to arise from a superposition of many incoherent
sources of gravitational waves, of either cosmological or astrophysical origin.
This background is a target for the current generation of ground-based
detectors. In this article we present the first joint search for a stochastic
background using data from the LIGO and Virgo interferometers.
In a frequency band of 600-1000 Hz, we obtained a $95\%$ upper limit on the
amplitude of 
$\Omega_{\rm GW}(f) = \Omega_3 \left( f/900 \mathrm{Hz}\right)^3$, of
$\Omega_3 < 0.33$, 
assuming a value of the Hubble parameter of
$h_{100}=0.72$. These new limits are a factor of 
seven better than the previous best in this frequency band.
\end{abstract}

\maketitle

\section{Introduction}
\label{sec:introduction}

A major science goal of current and future generations of gravitational-wave detectors is the detection of a stochastic gravitational wave
background (SGWB) -- a superposition of unresolvable gravitational-wave
signals of astrophysical and/or cosmological origin. An astrophysical
background is expected to be comprised of signals originating from
astrophysical objects, for example 
binary neutron stars \cite{Regimbau:2005tv},
spinning neutron stars \cite{Regimbau:2001kx}, 
magnetars \cite{Regimbau:2005ey} or core-collapse supernovae \cite{Ferrari-et-al:1999}.
A cosmological background is expected to be generated by various
physical processes in the early universe \cite{Grishchuk:1976} and, as gravitational waves are so 
weakly interacting, to be essentially unattenuated since then. We expect 
that gravitational waves would decouple much earlier than other radiation, 
so a cosmological background would carry the earliest information accessible 
about the very early universe \cite{magg}. There are various
production mechanisms from which we might expect cosmological gravitational waves
including cosmic strings \cite{Siemens:2006yp}, 
amplification of vacuum fluctuations following inflation \cite{BarKana:1994bu,Starobinsky:1979ty}, 
pre-Big-Bang models \cite{Brustein:1995ah,Mandic:2005bd}, or the electroweak phase transition \cite{Apreda:2001us}.

Whatever the production mechanism of a SGWB,
 the signal is usually described in terms of the dimensionless quantity,
\begin{equation}
\Omega_{\rm GW}(f) = \frac{f}{\rho_c} \; \frac{d \rho_{\rm GW}}{df}\,,
\end{equation}
where $d\rho_{{\rm {GW}}}$ is the energy density of gravitational
radiation contained in the frequency range $f$ to $f+df$
and $\rho_c$ is the critical energy density of
the universe \cite{Allen:1997ad}.
As a SGWB signal is expected to be much smaller than 
current detector noise, and because we assume both the detector noise and 
the signal to be Gaussian random variables, it is not feasible to distinguish
the two in a single interferometer.
We must therefore search for the
SGWB using two or more interferometers.
The optimal method is to cross-correlate the strain
data from a pair, or several pairs of detectors \cite{Grishchuk:1976,Allen:1997ad}. 
In recent years, several interferometric gravitational wave detectors
have been in operation in the USA
and Europe. At the time that the data analysed in this paper were taken, 
five interferometers were in operation. 
Two LIGO interferometers were located at the same site in Hanford, WA,
one with 4km arms and one with 2km arms (referred to as H1 and H2 respectively).
In addition, one LIGO 4km interferometer, L1, was located in 
Livingston, LA \cite{LIGO}. The Virgo interferometer, V1, with 3km arms was 
located near Pisa, Italy \cite{Virgo}  
and GEO600, with 600m arms, was located near Hannover, Germany \cite{GEO600}. 
LIGO carried out its fifth
science run, along with GEO600, 
between 5th November 2005 and 30th September 2007. They were joined from 18th May 2007
by Virgo, carrying out its first science run.
In this paper we present a joint analysis of the data taken by the LIGO
and Virgo detectors 
during these periods,
in the frequency range 600-1000 Hz.
This is the first search for a SGWB using data from both
LIGO and Virgo interferometers, and the first using multiple baselines.
Previous searches using the LIGO interferometers used just one baseline. The most sensitive direct  limit obtained so far used the three LIGO interferometers, but as the two Hanford interferometers were colocated this involved
just one baseline \cite{Abbott:2009ws}. The most recent upper limit
in frequency band studied in this paper was obtained using data from the LIGO Livingston interferometer and the 
ALLEGRO bar detector, which were colocated for the duration of the analysis \cite{Abbott:2007wd}. 
The addition of Virgo to the LIGO interferometers adds two further
baselines, for which the frequency dependence of the sensitivity varies
differently.
 The frequency range used in this paper was chosen because the addition of
Virgo data was expected to most improve the sensitivity at these high 
frequencies. This is due in part to the relative orientation and 
separation of the LIGO and Virgo interferometers, and in part
to the fact that the Virgo sensitivity is
closest to the LIGO sensitivity at these frequencies. 
The GEO600 interferometer was not included in this analysis
as the strain sensitivity at these frequencies was insufficient to significantly
improve the sensitivity of the search.

The structure of this paper is as follows. In Section \ref{sec:analysis}
we describe the method used to analyse the data. 
In Section \ref{sec:results} we present the results
of the analysis of data from the LIGO and Virgo interferometers. 
We describe 
validation of the results using software 
injections in Section \ref{sec:validation}.
In Section \ref{sec:compare} we compare our results to those of previous
experiments and in 
Section \ref{sec:conclusion} we summarise our conclusions.

\section{Analysis Method}
\label{sec:analysis}

The output of an interferometer is assumed to be the sum of 
instrumental noise and a stochastic background signal,
\begin{equation}
s(t) = n(t) + h(t).
\end{equation}
The gravitational wave signal has a power spectrum, $S_{\rm GW}(f)$,
which is related to $\Omega_{\rm GW}(f)$ by \cite{Abbott:2006zx}
\begin{equation}
S_{\rm GW}(f) = \frac{3 H_0^2}{10 \pi^2} \; 
\frac{\Omega_{\rm GW}(f)}{f^3} \; .
\label{eq:Sgw}
\end{equation}
%
Our signal model is a power law spectrum,
\begin{equation}
\Omega_{\mathrm{GW}}(f) = \Omega_{\alpha} \left( \frac{f}{f_R }\right)^\alpha,
\end{equation}
where $\alpha$ is the spectral index, and $f_R$ a reference frequency, such
that $\Omega_{\alpha} = \Omega_{\rm GW}(f_R)$. 
For this analysis we create a filter using a model which corresponds to a white
strain amplitude spectrum and choose a reference frequency of 900 Hz, such that
\begin{equation}
\Omega_{\mathrm{GW}}(f) = \Omega_3 \left( \frac{f}{900 \mathrm{Hz} }\right)^3.
\end{equation}
We choose this spectrum as it is expected that some 
astrophysical backgrounds will
have a rising $\Omega_{\rm GW}(f)$ spectrum in the frequency band we are investigating 
\cite{Regimbau:2001kx,Regimbau:2005ey,Ferrari-et-al:1999}. In fact, 
different models predict different values of the spectral index 
$\alpha$ in our frequency band, so we quote upper limits for several 
values.

For a pair of detectors, with interferometers labelled by $i$ and $j$, 
we calculate the cross-correlation statistic
in the frequency domain
\begin{eqnarray}
\hat{Y} & = & \int_{-\infty}^{+\infty } df \; Y(f)  \\
& = & \int_{-\infty }^{+\infty } df \int_{-\infty }^{+\infty } df' \;
\delta_T (f-f') 
\; \tilde{s}^\star_i(f) \; \tilde{s}_j(f') \; \tilde{Q}_{ij}(f')\,,\nonumber
\label{eq:ccstat}
\end{eqnarray}
where
$\tilde{s}_i(f)$ and $\tilde{s}_j(f)$ are the Fourier transforms of the
strain time-series of two interferometers, $\tilde{Q}_{ij}(f)$ is
a filter function and  $\delta_T$ is a finite-time approximation to the Dirac delta function, \cite{Allen:1997ad}
\begin{equation}
\delta_T(f):=\int_{-T/2}^{T/2} df e^{-i2\pi f t} = \frac{\sin{\left( \pi f T \right)}}{\pi f}.
\end{equation}
We assume the detector noise is Gaussian, stationary, 
uncorrelated between the two interferometers and much larger than the signal.
Under these assumptions, the variance of the estimator $\hat{Y}$ is
\begin{eqnarray}
\sigma_Y^2 & = & \int_0^{+\infty} df \; \sigma_Y^2(f) \nonumber \\
& \approx & \frac{T}{2} \int_0^{+\infty} df P_i(f) P_j(f)
\mid \tilde{Q}_{ij}(f) \mid^2\,,
\label{eq:sigma}
\end{eqnarray}
where $P_i(f)$ is the one-sided power spectral density of interferometer $i$
and $T$ is the integration time. 
By maximizing the expected signal-to-noise ratio (SNR) for a chosen model of 
$\Omega_{\rm GW}(f)$, we find the optimal filter function, 
\begin{equation}
\label{eq:optfilt}
\tilde{Q}_{ij}(f) = \mathcal{N} \; \frac{\gamma_{ij}(f) 
f^{\alpha-3}}{f_R^{\alpha}P_i(f) P_j(f)} \; ,
\end{equation}
where 
$\gamma_{ij}(f)$ is the overlap
reduction function (ORF) of the two interferometers and 
$\mathcal{N}$ is a normalisation factor. 
We choose the normalisation such that the cross-correlation
statistic is an estimator of $\Omega_{\alpha}$, with expectation value $\langle \hat{Y} \rangle = \Omega_{\alpha}$. It follows that the
normalisation is
\begin{equation}
\mathcal{N} = \frac{f_R^{2\alpha}}{2T}\left( \frac{10\pi^2}{3H_0^2}\right)
\left[ \int_0^{\infty} df \frac{f^{2\alpha-6}\gamma_{ij}^2(f)}{P_i(f) P_j(f)} \right]^{-1}.
\label{eq:norm}
\end{equation}
Using this filter function and normalisation gives an optimal SNR of \cite{Allen:1997ad}
\begin{equation}
\mathrm{SNR}\approx \frac{3 H_0^2}{10 \pi^2}\sqrt{2 T} \left[ \int_{0}^{\infty} 
\frac{\gamma_{ij}^2(f)\Omega^2_{\alpha}(f)}{f^6 P_1(f) P_2(f)}\right]^{1/2}.
\end{equation}

\begin{figure}
\includegraphics[width=8cm]{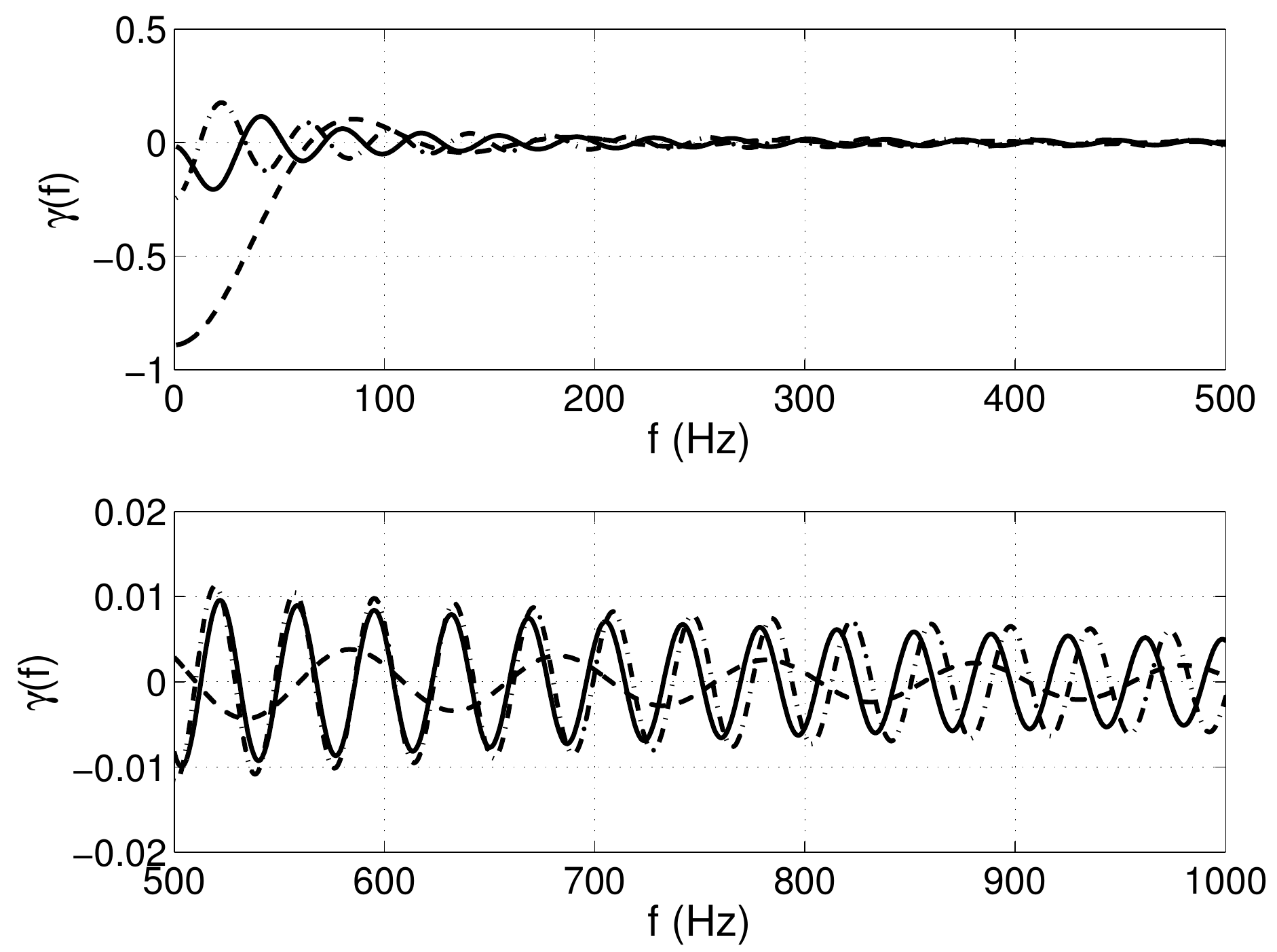}
\caption[Plot of the overlap reduction function for several detector pairs]
{Plot of the overlap reduction function (ORF) for the pairs of sites used in 
this analysis. The dashed curve is the ORF for the two LIGO sites (HL), the 
solid curve is for the Hanford-Virgo sites (HV) and the dashed-dotted curve is
for Livingston-Virgo (LV). We see that the LIGO 
orientations
have been optimized for low frequency searches, around 10--100 Hz. However,
this ORF falls off rapidly with frequency, such that at frequencies over
$\sim 500$ Hz, the amplitude of the ORF of the HL pair is smaller than 
that of the 
Virgo pairs. The LV and HV overlap reduction functions oscillate
more with frequency, but fall off more slowly,
due to the larger light-travel time between 
the USA and Europe.\label{fig:orf}}
\end{figure}

The ORF encodes the separation and 
orientations of the detectors and is defined as \cite{Flanagan:1993,Allen:1997ad}
\begin{equation}
\gamma_{ij}(f) := \frac{5}{8\pi} \sum_A \int_{S^2} d\hat{\Omega} 
e^{i 2 \pi f \hat{\Omega} \cdot \Delta \vec{x}/c}
F_i^A(\hat{\Omega}) F_j^A(\hat{\Omega}),
\end{equation}
where $\hat{\Omega}$ is a unit vector specifying a direction on the two-sphere,
$\Delta\vec{x}=\vec{x}_i-\vec{x}_j$ is the separation of the two 
interferometers and 
\begin{equation}
F_i^A(\hat{\Omega}) = e_{ab}^A(\hat{\Omega}) d_i^{ab}
\end{equation}
is the response of the $i$th detector to the $A=+,\times$ polarisation,
where $e_{ab}^A$ are the transverse traceless polarisation tensors.
The geometry of each interferometer is described by a response tensor,
\begin{equation}
d^{ab} = \frac{1}{2}(\hat{x}^a\hat{x}^b - \hat{y}^a\hat{y}^b),
\end{equation}
which is constructed from the two unit vectors that point along
the arms of the interferometer, $\hat{x}$ and $\hat{y}$ \cite{Flanagan:1993,Christensen:1992}.
At zero frequency, the ORF is determined solely by the 
relative orientations of the two interferometers. The LIGO interferometers 
are oriented in such a way as to maximize the amplitude of the ORF at 
low frequency, while the relative orientations 
of the LIGO-Virgo pairs are poor. Thus at low frequency the amplitude 
of the ORF between the Hanford and Livingston interferometers, 
$\gamma_{HL}(f)$, is larger than that of the overlap between Virgo and
any of the LIGO interferometers, $\gamma_{HV}(f)$ or $\gamma_{LV}(f)$
(note that the `HL'
and `HV' overlap reduction functions hold for both H1 and H2 as they 
are colocated).
However, at high frequency the ORF behaves as a sinc function of the 
frequency multiplied by the light travel time between the interferometers.
As the LIGO interferometers are closer to each other than to Virgo, 
their ORF $\gamma_{HL}(f)$ oscillates less, but decays more
rapidly with frequency than the the ORFs of the LIGO-Virgo pairs. 
Fig.~\ref{fig:orf} shows the ORFs 
between the LIGO Hanford, LIGO Livingston and Virgo sites.

We 
define the ``sensitivity integrand'', $\mathcal{I}(f)$, by inserting 
Eq.~\ref{eq:optfilt} and Eq.~\ref{eq:norm} into Eq.~\ref{eq:sigma}, giving
\begin{eqnarray}
\sigma_Y^{2} & = &
\left(\int_{0}^{\infty} \mathcal{I}(f) df \right)^{-1}\nonumber\\
 & = & \frac{f_R^{2\alpha}}{8T} \left( \frac{10\pi^2}{3H_0^2}\right)^2
\left(\int_{0}^{\infty} df  
\frac{\gamma_{ij}^2(f)f^{2\alpha-6}}{P_i(f)P_j(f)}\right)^{-1}.
\label{eq:sensint}
\end{eqnarray}
This demonstrates the contribution to the inverse of the variance at each 
frequency. 
The sensitivity of each pair is dependent on the noise power spectra of
the two interferometers, as well as the observing geometry, 
described by $\gamma_{ij}(f)$. 
For interferometers operating at design sensitivity, this means that 
for frequencies above $\sim 200$ Hz the LIGO-Virgo pairs make the 
dominant contribution to the 
sensitivity \cite{CellaEtAl:2007}.
During its first science run Virgo was closest to design sensitivity 
at frequencies above several hundred Hz, which informed our decision to use the 
600-1000 Hz band.

The procedure by which we analysed the data is as follows. For each
pair of interferometers, labelled by $I$, the coincident data 
were divided into segments, labelled by $J$, of length 
$T=60$s. 
The data 
from each segment are Hann windowed in order to minimize 
spectral leakage. In order not to reduce the effective observation time, the
segments are therefore overlapped by $50\%$.
For each segment, the data from both interferometers
were Fourier transformed then coarse-grained to a resolution of 0.25 Hz. The data from the adjacent
segments were then used to calculate power spectral densities (PSDs) with Welch's method.
The Fourier transformed data and the PSDs were used to 
calculate the estimator on $\Omega_{3}$, 
$\hat{Y}_{IJ}$, and its standard deviation, $\sigma_{IJ}$.
 For each pair, the 
results from all segments were 
optimally combined by performing a weighted average 
(with weights $1/\sigma_{IJ}^2$), taking into account the correlations 
that were introduced by the overlapping segments \cite{LazzariniRomano:2004}. The weighted average for each pair, $\hat{Y}_I$, 
has an associated standard deviation $\sigma_I$, 
also calculated by combining the standard deviations from each segment (note that $\sigma_{I}$ is
the equivalent of $\sigma_Y$ (from Eq.~\ref{eq:sigma}) for each pair, $I$, but we have dropped the $Y$ subscript 
to simplify the notation).

\subsection{Data Quality}

Data quality cuts were made to eliminate data that was too noisy or 
non-stationary, or that had correlated noise between detectors.
Time segments that were known to 
contain large noise transients 
in one interferometer were removed from the analysis.
We also excluded times when the digitizers were saturated, times with 
particularly high noise and times when the calibration was unreliable. This
also involved excluding the last thirty seconds before the loss of lock
in the interferometers,
as they are known to have an increase in noise in this period.
Additionally, we ensured that the data were approximately stationary over a period of
three minutes, as the PSD estimates, $P_{i}(f)$, used in calculating the
optimal filter and standard deviation in each segment are obtained from data in
the immediately adjacent segments.
This was achieved by calculating a measure of stationarity,
\begin{equation}
\Delta\sigma_{IJ} = \frac{|\sigma_{IJ} - \sigma_{IJ}'|}{\sigma_{IJ}},
\end{equation}
for each segment,
where $\sigma_{IJ}$ was calculated (following Eq.~\ref{eq:sigma}) using 
the PSDs estimated from the adjacent segments, and $\sigma_{IJ}'$ was 
calculated using the PSDs estimated using data from the segment itself. 
To ensure stationarity, we set a threshold value, $\zeta$, 
and
all segments with values of $\Delta\sigma_{IJ} > \zeta$
are discarded. The threshold was tuned by analysing the data with
unphysical time offsets between the interferometers; a value of $\zeta=0.1$
as this ensures that the remaining data are Gaussian.

In order to exclude correlations between the instruments
caused by environmental factors we excluded certain frequencies from our
analysis. The frequency bins to be removed were identified in two ways. Some correlations
between the interferometers were known to exist a priori, e.g.~there are 
correlations at multiples of 60 Hz between the interferometers located in the 
USA due to the frequency of the power supply \cite{Abbott:2006zx}. These were removed from the 
analysis, but in order to ensure that all 
coherent bins were identified, we also calculated the coherence, 
\begin{equation}
\Gamma(f) = \frac{\left|\tilde{s}_1^\star(f)\tilde{s}_2(f)\right|^2}{P_1(f)P_2(f)},
\end{equation}
which is the ratio of the cross-spectrum to the product of the two
power spectral densities, averaged over the whole run. This value was 
calculated first at a resolution of 0.1 Hz, then at 1 mHz to investigate in more detail the 
frequency distribution of the coherence. Several frequencies showed
excess coherence; some had been identified a priori but two had not, so these 
were also removed from the analysis. The calculations of the power spectra and the cross-correlation
were carried out at a resolution of 0.25 Hz, so we removed the corresponding 0.25 Hz bin from our analysis.
Excess coherence was defined as coherence exceeding a threshold of 
$\Gamma(f)=5\times 10^{-3}$. 
This threshold was also chosen after analysing the data with unphysical time
offsets.
 The excluded bins for each interferometer can be seen in Table \ref{t:notch}.

\begin{table}[h]
\centering
\begin{tabular}{|c||l|l|}
\hline
IFO 	&  \multicolumn{2}{l|}{Notched frequencies (Hz)}\\
\hline
H1	& 786.25	& Harmonic of calibration line	\\
	& 961		& Timing diagnostic line	\\
\hline
H2	& 640 		& Excess noise	\\
	& 814.5	& Harmonic of calibration line	\\
	& 961 		& Timing diagnostic line	\\
\hline
L1	& 793.5	& Harmonic of calibration line	\\
	& 961 		& Timing diagnostic line	\\
\hline
V1	& 706		&	\\
	& 710		& Harmonics of	\\
	& 714		& calibration lines	\\
	& 718 		&	\\
\hline
\end{tabular}

\caption[Table of frequency bins excluded from the analysis.]
{Table of the the frequency bins excluded from the analysis for each 
interferometer. The bins at 640 Hz and 961 Hz were identified using coherence tests, 
while the others were excluded a priori. Also excluded were harmonics of the power line frequency
at multiples of 60 Hz for the LIGO 
detectors and 
multiples of 50 Hz for Virgo.
Each excluded bin is centred at the frequency listed above and has a width of 0.25 Hz.
\label{t:notch}}
\end{table}

\subsection{Timing accuracy}
\label{sec:timing}

In order to be sure that the cross-correlation is a measure of the
gravitational-wave signal present in both detectors in a pair,
we must be sure that the data collected in both detectors are truly 
coincident. 
Calibration studies were carried out to determine the timing offset, if any,
 between the detectors and to estimate the error on this offset. These
studies are described in more detail in reference \cite{timing}, but we
summarize them here.

The output of each interferometer is recorded at a rate of
16384 Hz. Each data point has an associated 
time-stamp and we need to ensure that data taken with identical time
stamps are indeed coincident measurements of the strain, to within the
calibration errors of the instruments. No offset between the instruments
was identified, but several possible sources of timing error were investigated.
First, approximations in our models of the interferometers can introduce
phase errors. 
For the measurement of strain, we model the interferometers using the
long-wavelength approximation (i.e.~we assume that 
the wavelengths of the gravitational waves that we 
measure are much longer than the arm-lengths of the interferometers).
We also make an approximation in the transfer function of the Fabry-Perot
cavity; the exact function has several poles or singularities, but we use an
approximation which includes only the lowest frequency pole \cite{Rakhmanov:2008}.
The errors that these two approximations introduce largely cancel, 
with a residual error of $\sim 2\mu\mathrm{s}$ or $\sim 1^\circ$ at 1kHz \cite{timing}. 

Secondly, there is some propagation time between strain manifesting in 
the detectors 
and the detector output being recorded in a frame file. This is well 
understood for all detectors and is accounted for (to within calibration 
errors) when the detector outputs are converted to strain.
The time-stamp associated with each data point is therefore taken to be
the GPS time at 
which the differential arm length occurs, to within calibration errors \cite{timing}.

Thirdly, the GPS time recorded at each site has some uncertainty. 
The timing precision of the GPS system is  
$\sim 30$ ns, which corresponds with the stated location accuracy of
$\sim 10$ m. Each site necessarily uses its own GPS receiver, so the relative
accuracy of these receivers has been checked, by taking a Virgo GPS receiver
to a LIGO site and comparing the outputs. The relative accuracy was found 
to be better than
$1 \mu \mathrm{s}$. The receivers have also been checked against 
Network Time Protocol (NTP) and were found to have no offset \cite{timing}. 
The total error in GPS timing is
 far smaller than the instrumental phase calibration errors in
the 600-1000 Hz frequency band (see Table \ref{t:calibration}).

These investigations concluded that the timing offset between the 
between the instruments is zero for all pairs, with errors on these values that are 
smaller than the error in the phase calibration of each 
instrument.  The phase calibration errors of the instruments are negligible in
this analysis as their inclusion would produce a smaller than $1\%$ change in the
results at this sensitivity, and therefore the relative timing error is negligible.

\subsection{Combination of multiple pairs}

We performed an analysis of all of the available data from LIGO's fifth
science run and Virgo's first science run. However, we excluded the H1-H2 pair 
as two 
instruments were built inside the same vacuum system, and so may
have significant amounts of correlated noise. 
There is an ongoing 
investigation into identifying and removing these
correlations \cite{Fotopoulos:2008}, and for the present analysis, we consider 
only the five remaining 
pairs. As described above, the output of each pair yields an estimator, $\hat{Y}_I$, with a
standard deviation, 
$\sigma_{I}$, where $I = 1\ldots5$ labels the detector pair.

Using the estimators $\hat{Y}_I$ and their associated error bars, $\sigma_I$,
we construct a Bayesian posterior probability density function (PDF) on $\Omega_3$.
Bayes theorem says that the posterior PDF of a set of unknown parameters, $\vec{\theta}$,
given a set of data, $\mathcal{D}$, is given by
\begin{equation}
p(\vec{\theta}|\mathcal{D}) = \frac{p(\vec{\theta})p(\mathcal{D}|\vec{\theta})}{p(\mathcal{D})},
\end{equation}
where $p(\vec{\theta})$ is the prior PDF on the unknown parameters -- representing the state of
knowledge before the experiment -- $p(\mathcal{D}|\vec{\theta})$ is the likelihood function --
representing the probability distribution of the data given particular values of the unknown parameters --
and $p(\mathcal{D})$ is a normalisation factor.
In this case, the unknown parameters, $\vec{\theta}$, are the value of $\Omega_3$ and the amplitude calibration factors of
the instruments, which will be discussed below. The data set, $\mathcal{D}$, is the set of five estimators, 
$\lbrace \hat{Y}_I\rbrace$, 
we obtain from the five pairs of instruments.

In forming this posterior, we must consider the errors in the 
calibration of the strain data obtained
by the interferometers. In the data from one interferometer, labelled by $i$,
there may be an  error on 
the calibration of both the amplitude and the phase, such that the value
we measure is
\begin{equation}
\tilde{s}_i(f) = e^{\Lambda_i + i\phi_i}\tilde{s}^t_i(f),
\end{equation}
where $\tilde{s}^t_i(f)$ is the ``true'' value that would be measured if the 
interferometer were perfectly calibrated. 
The phase calibration errors given in Table \ref{t:calibration} are negligible, and the
studies described in Section \ref{sec:timing} have shown that there is no significant
relative timing error between the interferometers, so
we can simply assume that $\phi_i=0$.
However, the amplitude calibration errors are not negligible, and
the calibration factors take the values $\Lambda_i=0\pm\epsilon_{\Lambda,i}$,
where $\epsilon_{\Lambda,i}$
are the fractional amplitude calibration errors of the instruments, which are quoted in Table 
\ref{t:calibration}.

\begin{table}[h]
\centering
\begin{tabular}{|c||c|c|}
\hline 
Instrument & Amplitude error ($\%$) & Phase error (deg) \\
\hline
H1	   & 10.2		    & 4.3		\\
\hline
H2	   & 10.3		    & 3.4		\\
\hline
L1	   & 13.4		    & 2.3		\\
\hline
V1	   & 6.0		    & 4.0		\\
\hline
\end{tabular}

\caption[Table of values of the calibration errors for each instrument used.]
{Table of values of the errors in the calibration of amplitude and phase 
for each of the LIGO \cite{Abadie:2010px} and Virgo \cite{Accadia:2010sk} 
instruments used in this analysis. The
errors are valid over the whole 600-1000 Hz band.
\label{t:calibration}}
\end{table}

The calibration factors combine such that the estimator for a 
pair $I$ is
\begin{equation}
\hat{Y}_I = e^{\Lambda_{I,1}+ \Lambda_{I,2}} \hat{Y}^t_I,
\end{equation}
where $\hat{Y}^t_I$ is the ``true'' value that would be measured with perfectly 
calibrated instruments and $\Lambda_{I,1}$ and $\Lambda_{I,2}$ are the calibration factors of the
two instruments in pair $I$.
The likelihood function for a single estimator is
is given by
\begin{equation}
p(\hat{Y}_I|\Omega_3,\sigma_I,\Lambda_{I}) = \frac{1}{\sigma_I\sqrt{2\pi}}
\exp{\left( -\frac{\left(\hat{Y}_I-e^{\Lambda_{I}}\Omega_3\right)^2}{2\sigma_I^2} \right)},
\label{eq:singlike}
\end{equation}
where we have used $\Lambda_I = \Lambda_{I,1}+\Lambda_{I,2}$.
The joint likelihood function on all the data is the product over all pairs
of Equation \ref{eq:singlike}
\begin{equation}
p(\lbrace \hat{Y}_I\rbrace|\Omega_3,\lbrace\sigma_I\rbrace,\lbrace{\Lambda_I}\rbrace) = \prod_{I=1}^{n_{\mathrm{pairs}}} p(\hat{Y}_I|\Omega_3, \sigma_I ,\Lambda_I).
\end{equation}

In order to form a posterior PDF, we define priors on the 
calibration factors of the individual interferometers, 
$\lbrace\Lambda_i\rbrace$. The calibration factors are assumed to be 
Gaussian distributed, with variance given by the square of the calibration
errors quoted in Table \ref{t:calibration}, such that
\begin{equation}
p(\lbrace \Lambda_i\rbrace|\lbrace\epsilon_{\Lambda,i}\rbrace) = \prod_{i=1}^{n_{\mathrm{IFO}}} 
\frac{1}{\epsilon_{\Lambda,i}\sqrt{2\pi}} \exp{\left( -\frac{\Lambda_i^2}{2\epsilon_{\Lambda,i}^2}\right)},
\end{equation}
where $n_{\mathrm{IFO}}$ is the number of interferometers we are using, in this case four.
The prior on $\Omega_3$ is a top hat function
\begin{equation}
p(\Omega_3) =\left\{
\begin{array}{ll}
\frac{1}{\Omega_{\mathrm{max}}} & \mathrm{for}\,\, 0\leq\Omega<\Omega_{\mathrm{max}} \\
0 & \mathrm{otherwise}
\end{array}
\right. .
\end{equation}
We choose a flat prior on $\Omega_3$ because, although there has been an analysis in this band previously,
it did not include data from the whole of the frequency band and an uninformative flat prior is conservative. We chose $\Omega_{\mathrm{max}}=10$, which is two orders of magnitude greater than the 
estimators and their standard deviations, such that the prior is essentially unconstrained.

We combine the prior and likelihood functions to give a posterior PDF
\begin{eqnarray}
p(\Omega_3,\lbrace \Lambda_i \rbrace|\lbrace \hat{Y}_I\rbrace ,\lbrace \sigma_I\rbrace,\lbrace \epsilon_{\Lambda,i}\rbrace) &=& 
p(\Omega_3) p(\lbrace \Lambda_i \rbrace|\lbrace \epsilon_{\Lambda,i}\rbrace) \ldots\\
&&\times p(\lbrace \hat{Y}_I \rbrace|\Omega_3,\lbrace \sigma_I\rbrace,\lbrace \Lambda_I\rbrace)\nonumber.
\end{eqnarray}
We marginalize this posterior analytically over all $\Lambda_i$ to give us a
posterior on $\Omega_3$ alone,
\begin{widetext}
\begin{eqnarray}
p(\Omega_3|\lbrace \hat{Y}_I\rbrace ,\lbrace \sigma_I \rbrace ,\lbrace\epsilon_{\Lambda,i}\rbrace) &=&
\int_{-\infty}^{\infty} d\Lambda_1 \int_{-\infty}^{\infty} d\Lambda_2 \ldots \int_{-\infty}^{\infty} d\Lambda_{n_{\mathrm{IFO}}}  p(\Omega_3,\lbrace \Lambda_i \rbrace|\lbrace \hat{Y}_I\rbrace ,\lbrace \sigma_I\rbrace,\lbrace \epsilon_{\Lambda,i}\rbrace).\label{e:posterior}
\end{eqnarray}
\end{widetext}
Using this posterior PDF
we calculate a $95\%$ probability interval, 
$(\Omega_{\mathrm{lower}},\Omega_{\mathrm{upper}})$ on $\Omega_3$. We calculate the values of 
$\Omega_{\rm lower}$ and $\Omega_{\rm upper}$ by finding the minimum-width interval that satisfies
\begin{equation}
\int_{\Omega_{\mathrm{lower}}}^{\Omega_{\mathrm{upper}}} 
p(\Omega_3|\lbrace \hat{Y}_I\rbrace ,\lbrace \sigma_I \rbrace ,\lbrace\epsilon_{\Lambda,i}\rbrace)  d\Omega_3= 0.95.
\end{equation}
If we find that $\Omega_{\mathrm{lower}}$ is equal to zero, then we have a null result, and
we can simply quote the upper limit, $\Omega_{\mathrm{upper}}$.

The optimal estimator, $\hat{Y}$, is given by the combination of $Y_I$, 
$\sigma_I$ and $\Lambda_I$ that maximises the likelihood, such that
\begin{equation}
\hat{Y} = \frac{\sum_I e^{\Lambda_I} \hat{Y}_I \sigma_I^{-2}}{\sum_I e^{2\Lambda_I}\sigma_I^{-2}}.
\end{equation}
It has a variance, $\sigma$, given by
\begin{equation}
\sigma^{-2} =\sum_I e^{2\Lambda_I}\sigma_I^{-2}.
\end{equation}
Under the assumption that the calibration factors $\Lambda_i$ are all equal to zero,
then the optimal way to combine the results from each pair is to perform a weighted 
average with weights $1/\sigma_I^2$ (equivalently to combining results from
multiple, uncorrelated, time segments) \cite{CellaEtAl:2007}
\begin{eqnarray}
\hat{Y} &= &\frac{\sum_I \hat{Y}_I\sigma_{I}^{-2}}{\sum_I \sigma_{I}^{-2}}\label{eq:Y}\\
\sigma^{-2} &=& \sum_{I}\sigma_{I}^{-2}\label{eq:sig}.
\end{eqnarray}
A combined sensitivity integrand can also be found by summing the integrands 
from each pair: \cite{CellaEtAl:2007}
\begin{equation}
\mathcal{I}(f) = \sum_I \mathcal{I}_I(f)
\end{equation}

\section{Results}
\label{sec:results}

We applied the analysis described in Section \ref{sec:analysis} to
all of the available data from the LIGO and Virgo interferometers between
November 2005 and September 2007
\footnote{
We initially analyzed only data from times after Virgo had begun taking data (May--September 2007).  This preliminary analysis resulted in a marginal signal with a false-alarm probability of p=2\%.  To follow up, we extended the analysis to include all available LIGO data, yielding the results shown here, which are consistent with the null hypothesis.
}
and obtained
estimators of $\Omega_3$ from each of five pairs, which are listed in Table 
\ref{t:results} along with their standard deviations. We also create the combined estimators and their standard deviations,
using Equations \ref{eq:Y} and \ref{eq:sig}, for the full network, and 
for the network including only the LIGO interferometers. 
We see that the addition of Virgo to the network reduces the size of the
standard deviation by $23\%$.

\begin{table}[h]
\centering
\centering
\begin{tabular}{|c||r@{$\pm$}l|c|}
\hline
Network & \multicolumn{2}{c|}{Estimator $\hat{Y}_I$} \\
\hline
H1L1 & 0.11 & 0.15 \\
H1V1 & 0.57 & 0.22 \\
H2L1 & -0.14 & 0.26 \\
H2V1 & -0.52 & 0.41 \\
L1V1 & 0.18 & 0.20 \\
\hline
LIGO & 0.05 & 0.13 \\
\hline
all & 0.15 & 0.10 \\
\hline
\end{tabular}

\caption[Table of values of $\hat{Y}_I$ obtained 
from the data.]
{Table of values of $\hat{Y}_I$, the estimator of $\Omega_3$, 
obtained by analysing
the data taken during LIGO's fifth science run and Virgo's first science run, 
over a frequency band of 600-1000 Hz, along with the standard deviation, $\sigma_I$, of each result.
\label{t:results}}
\end{table}

\begin{figure}[h]
\centering
\includegraphics[width=8cm]{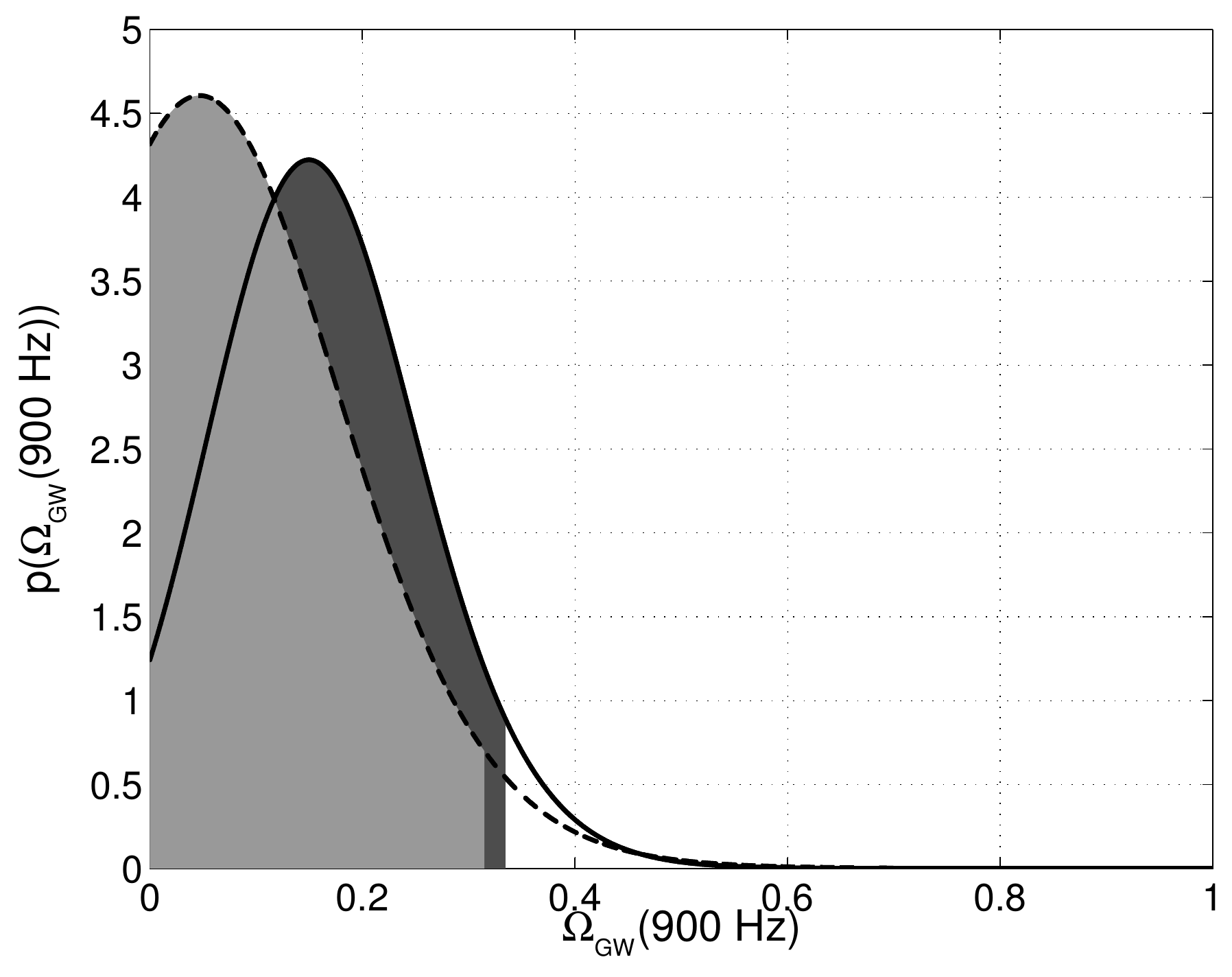}
\caption{Posterior PDFs on $\Omega_3$.
The dashed
line shows the posterior PDF obtained using just the LIGO detectors,
the solid line shows the PDF obtained using LIGO and Virgo detectors. 
The filled areas show the $95\%$ probability intervals.\label{f:posteriors}}
\end{figure}

\begin{figure}[h]
\centering
\includegraphics[width=8cm]{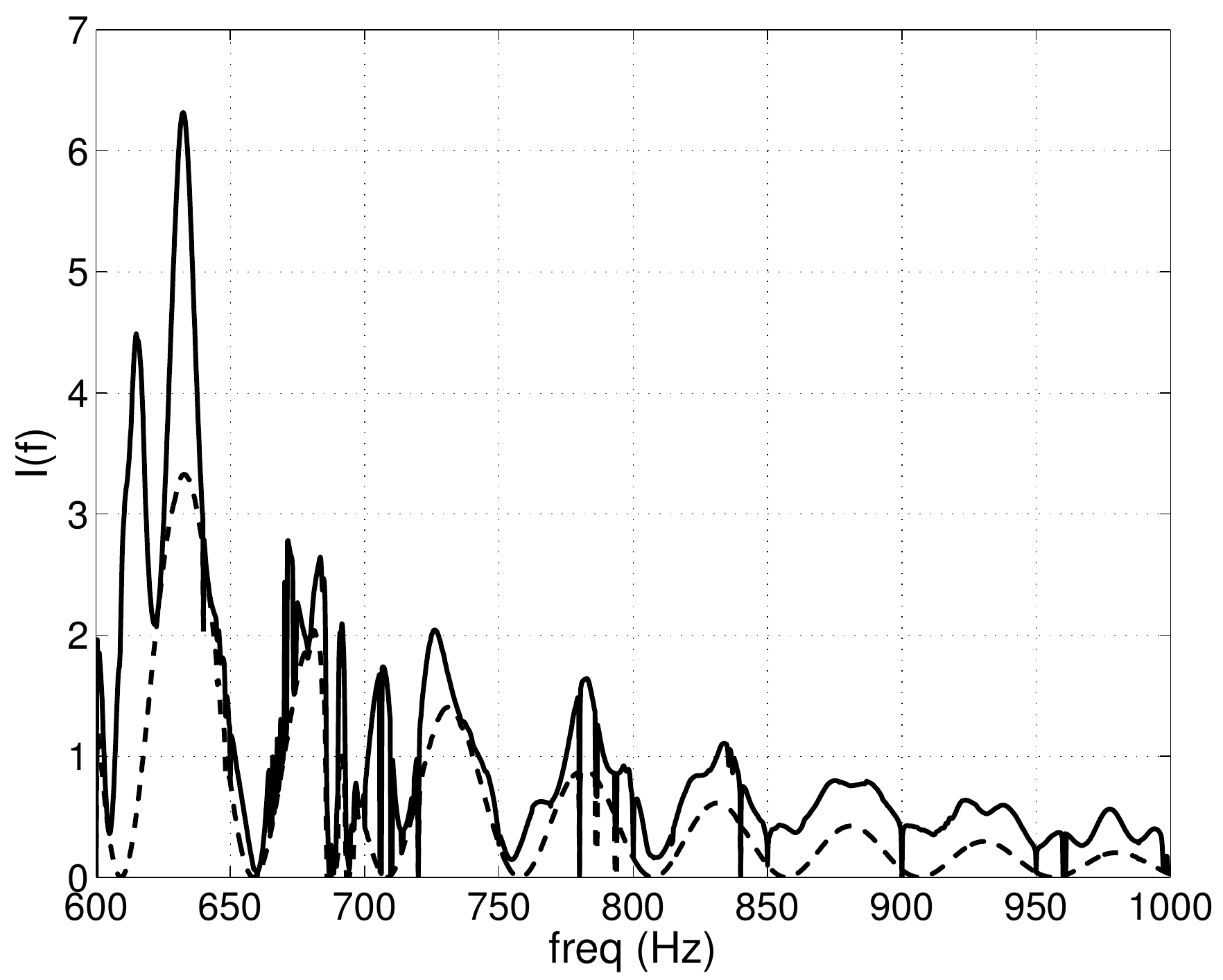}
\caption{Sensitivity integrands for the LIGO only result (dashed)
and for the full LIGO-Virgo result (solid). We can see that the sensitivity
is increased across the band by the addition of the Virgo interferometer to 
the search. The vertical lines correspond to frequency bins removed from
the search.\label{f:sensint}}
\end{figure}

Using the posterior PDF defined in Equation \ref{e:posterior}
and the calibration errors in Table \ref{t:calibration}
we found a $95\%$ upper limit 
of $\Omega_3 < 0.33$, assuming the Hubble constant to be
$h_{100} = 0.72$ \cite{Bennett:2003bz},
while using only the LIGO instruments obtained an upper limit of
$\Omega_3 < 0.31$. Both of the lower limits were zero.
The posterior PDFs obtained by the search are shown in Figure 
\ref{f:posteriors}, while the sensitivity integrands, which show the 
contribution to the sensitivity of the search from each frequency bin, are shown
in Figure \ref{f:sensint}.
The upper limit 
corresponds to a strain sensitivity of 
$8.5\times 10^{-24} \mathrm{Hz}^{-1/2}$ using just the LIGO interferometers,
or $8.7\times 10^{-24} \mathrm{Hz}^{-1/2}$ using both LIGO and Virgo.  
The LIGO-only upper limit is, in fact, lower than the upper limit using the 
whole data set, even though the sensitivity of the combined LIGO-Virgo
analysis is better. This is not surprising because the addition of Virgo
also increases the value of the estimator. 
The estimator will usually lie somewhere between 0 and 2 $\sigma$ -- in this case, the LIGO-only estimator 
was in the lower part of that range while the LIGO-Virgo estimator was not, but the
two results are entirely consistent with each other. 
When we add Virgo, the likelihood
excludes more of the parameter space below $\Omega_3=0$, but this
is a region we already exclude by setting the priors.
Monte-Carlo simulations show that, in the absence of a signal, 
the probability of the combined LIGO-Virgo upper limit being at least this much larger than the 
LIGO-only upper limit is $4.3\%$. This probability is not so small as to indicate a non-null result and 
we therefore conclude that the LIGO-Virgo upper limit is larger due to statistical fluctuations.

\begin{figure}[h]
\centering
\includegraphics[width=8cm]{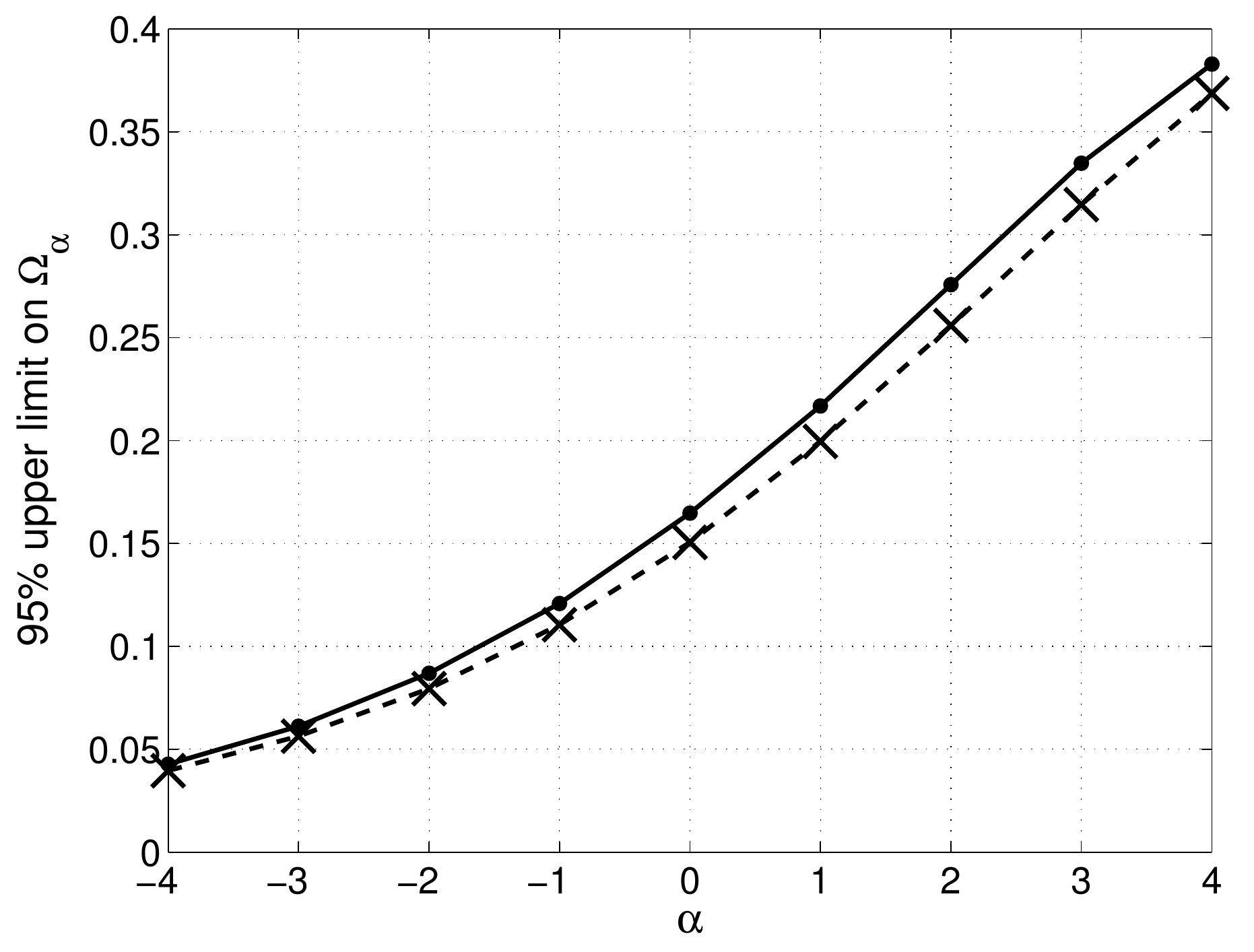}
\caption{95\% probability intervals on $\Omega_{\alpha}$, calculated using different
values of $\alpha$. These upper limits were all calculated using the same data,
with a band width of 600-1000 Hz and a reference frequency of 900 Hz. The 
dashed line shows the upper limit calculated using the LIGO interferometers
only, while the solid line shows the upper limits calculated using all
of the available data\label{f:alphas}. The lower limits were all zero.}
\end{figure}

We also used the same data to calculate the $95\%$ probability 
intervals for gravitational wave spectra with spectral indices 
ranging over $-4\leq \alpha \leq 4$, which correspond with different models
of possible backgrounds in our frequency band. For example, a background 
of magnetar signals would be expected to have a spectral index of $\alpha=4$ \cite{Regimbau:2005ey}.
Figure \ref{f:alphas} shows the values of these upper limits. Note that 
they were all calculated using a reference frequency of 900 Hz, and
Hubble parameter $h_{100}=0.72$.

\section{Validation of results}
\label{sec:validation}

\begin{figure}
\includegraphics[width=8cm]{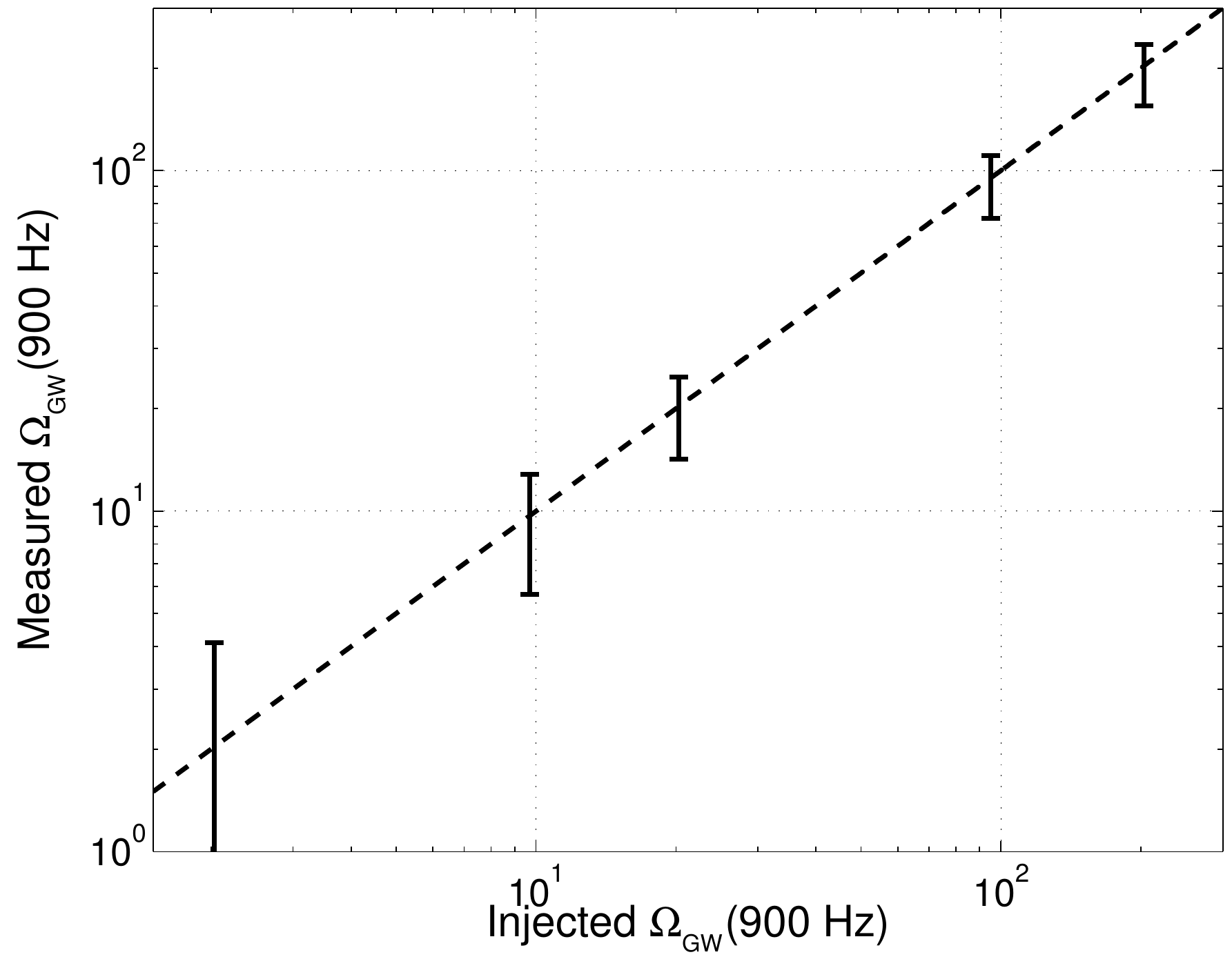}
\caption[The recovered values of $\Omega_3$ for three
software injections.]{Plot of the recovered values of 
	$\Omega_3$ for five software
	injections. The error bars show the $95\%$ probability intervals. 
	The
	quietest injection had a lower limit equal to zero. 
Note that all
	analyses excluded H2.
	Each injection used the same data, with the simulated signal scaled to
 	different amplitudes.
	\label{f:SWinj_err}}
\end{figure}

In order to test our analysis pipeline, we created simulated signals and
used software to add them to the data that had been taken during the
first week of Virgo's first science run  (this week was then excluded from the
full analysis). 
We generated frame files containing a simulated isotropic stochastic 
background, with $\Omega_{\rm GW}(f) \propto f^3$. We were then able to 
scale this signal to several values of $\Omega_3$
and add it to the data taken from the instruments. We did not include
H2 in this analysis, but used only H1, L1 and V1.
Table \ref{t:SWinjAmp} shows the injected values of 
$\Omega_3$ and the recovered values and associated
standard deviations, along with the SNR of the signal in the H1V1 pair. 
The recovered $95\%$ probability intervals of the injections can be seen in Figure
\ref{f:SWinj_err}. The intervals all contain the injected value of $\Omega_3$.

It should be noted that, in order to 
have detectable signals in this short amount of data, the larger 
injections are no longer in  
the small signal limit. We usually make two assumptions based on this limit. The first
is the approximation in Eq.~\ref{eq:sigma}, which only holds if the 
signal is much smaller than the noise, as we are ignoring terms that are
first and second order in $\Omega_{\mathrm{GW}}(f)$ \cite{Allen:1997ad}. 
The second assumption enters into the calculation 
of the noise PSDs, $P_i(f)$. We calculate these directly from the data, 
as in the 
small-signal limit we can assume that 
$\langle |\tilde{s}_i(f)|^2 \rangle \approx \langle |\tilde{n}_i(f)|^2 \rangle$.The first assumption causes an over-estimation of the standard deviation, while the
second causes our ``optimal'' filter to no longer be quite optimal. 
If we ignore these assumptions, we will underestimate the theoretical error bar, $\sigma_Y$, and
the width of the posterior PDFs. However, we still find $95\%$ probability intervals that are consistent
with the injected signals.

\begin{table}[h]
\centering
\centering
\begin{tabular}{|c||r@{$\pm$}l|r@{, }l|c|}
\hline
Injected $\Omega_3$ & \multicolumn{2}{c|}{Estimator $\hat{Y}$} & \multicolumn{2}{c|}{95\% probability interval} & SNR in H1V1\\ 
\hline
2.0 & 1.8 & 1.3 & (0.0 & 4.1) & 1.3 \\ 
9.7 & 9.1 & 1.5 & (5.7 & 12.8) & 6.3 \\ 
20.2 & 19.3 & 1.8 & (14.2 & 24.8) & 13.3 \\ 
95.1 & 91.1 & 3.7 & (72.3 & 110.6) & 62.3 \\ 
203.1 & 194.1 & 6.2 & (154.9 & 234.3) & 133.1 \\ 
\hline
\end{tabular}

\caption[Table of values of $\Omega$ for software injections.]
{Table of values of $\Omega_3$ for software injections, along with the
recovered values, the $95\%$ probability interval and the
expected SNR of each injection in the H1V1 pair.
Note that the
standard deviations presented in this table  are
underestimated, as the injections are not in the 
small signal limit, however we still recover the signals within the
$95\%$ probability intervals.
\label{t:SWinjAmp}}
\end{table}

\section{Comparison with other results}
\label{sec:compare}
\begin{figure}
\includegraphics[width=8cm]{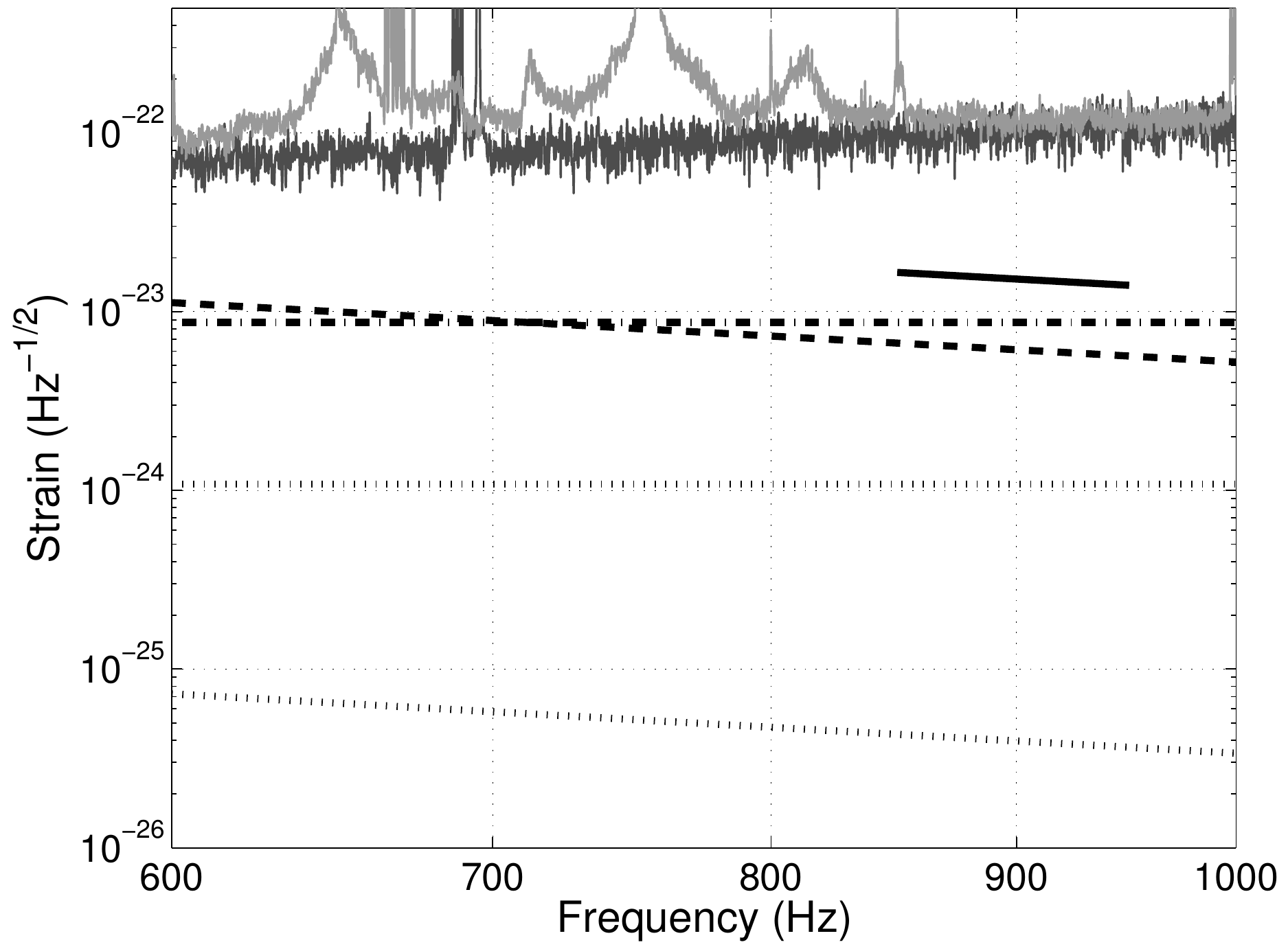}
\caption[Comparison of the strain sensitivity of two searches for a 
stochastic background of gravitational waves.]
{Comparison of the strain sensitivity of two searches for an isotropic 
stochastic background of gravitational waves. The two solid grey lines show  
 strain sensitivity of the Hanford 4km interferometer (dark grey)
and the Virgo interferometer (light grey), these spectra were obtained by 
averaging over the data analysed in this paper.
The dot-dashed line shows the main result of
this paper, the search for a SGWB with
$\Omega_{\rm GW}(f)\propto f^3$, which is white in strain amplitude, and corresponds to an upper limit 
of $\Omega_3 < 0.33$. The dashed line shows the
result of the same search, but for constant $\Omega_{\rm GW}(f)$, and 
corresponds to an upper limit of $\Omega_0<0.16$.
The solid black line shows the 
strain sensitivity of the LIGO-ALLEGRO search, which corresponds to an 
upper limit of $\Omega_0<1.02$ and was calculated over a frequency range
of $850\,\mathrm{Hz} \leq f \leq 950\,\mathrm{Hz}$ \cite{Abbott:2007wd}.
The two dotted lines show the extrapolation of the spectra obtained by the 
analysis of LIGO data in the frequency band  $40\,\mathrm{Hz} \leq f \leq 500\,\mathrm{Hz}$. 
The lower dotted line corresponds to a $95\%$ upper limit of $\Omega_0<6.9\times 10^{-6}$,
while the upper dotted line corresponds to an upper limit of $\Omega_3<0.0052$ at a reference 
frequency of 900 Hz.
\label{fig:strains}}
\end{figure}

The previous most sensitive direct upper limit in this frequency band was
$\Omega_{\rm GW}(f)<1.02$, obtained by the
joint analysis of data from the LIGO Livingston interferometer and the ALLEGRO
bar detector over a frequency band of
$850\,\mathrm{Hz} \leq f \leq 950\,\mathrm{Hz}$ \cite{Abbott:2007wd}. This result
was obtained using a constant $\Omega_{\rm GW}(f) = \Omega_0$, so should
be compared with our upper limit for $\alpha=0$. As can be seen in Figure
\ref{f:alphas}, our $95\%$ upper limit for $\alpha = 0$ is 
$\Omega_0 < 0.16$ using all the available data, or $\Omega_0 < 0.15$ using 
just the LIGO interferometers, therefore
our result has improved on the sensitivity of the LIGO-ALLEGRO result by a factor of 
$\approx 7$.
The comparative strain 
sensitivity of the upper limits of the current search and the LIGO-ALLEGRO search 
can be seen in Figure \ref{fig:strains}.  

The previous most sensitive direct limit at any frequency was the
analysis of data from the three LIGO detectors in the fifth science
run \cite{Abbott:2009ws}. The analysis was carried out using the same data
as the analysis presented in this paper, but was restricted
to the frequency band  $40\,\mathrm{Hz} \leq f \leq 500\,\mathrm{Hz}$. This included the most sensitive
frequency band of the three detectors. The $95\%$ upper limit on
$\Omega_0$ in this band was given as $6.9\times 10^{-6}$, which 
is a factor of $2\times 10^{4}$ times smaller than our upper limit.
They also found an upper limit on $\Omega_3$ of $7.1\times 10^{-6}$.
In order to compare that to our upper limit on $\Omega_3$, we
must extend the spectrum to the frequency band analysed in this paper. 
The  $40\,\mathrm{Hz} \leq f \leq 500\,\mathrm{Hz}$ upper limit would correspond to an upper limit at 900 Hz 
of $\Omega_3 < 0.0052$, which is a factor of $\approx 60$ smaller than
the upper limit presented in this paper. The search at lower frequencies is
significantly more sensitive and we 
would expect that in the advanced detector era the combined analysis of
LIGO and Virgo detectors at low frequencies will improve even further 
on the previously published upper limits.

We can also compare our results with indirect upper limits on the
stochastic gravitational wave background. In this band, the most stringent constraints
come from Big Bang nucleosynthesis (BBN) 
and measurements of the cosmic
microwave background (CMB).
The BBN bound constrains the
integrated energy density of gravitational waves
over frequencies above $10^{-10}$ Hz, 
based on observations of different relative abundances of light 
nuclei today. The BBN upper limit is \cite{magg}
\begin{equation}
\int \Omega_{\rm GW}(f) d\left( \ln{f}\right) < 1.1 \times 10^{-5} (N_{\nu}-3),
\end{equation}
where $N_{\nu}$ is the effective number of neutrino species at the time of
BBN. The most recent limit on this number is $(N_{\nu}-3)<4.6\times 10^{-2}$, 
given by measurements of 
the relative abundances of light elements and taking into account
neutrino oscillations \cite{Iocco:2008va}.
The CMB limit also constrains the integrated gravitational wave energy density, and 
is obtained from the observed CMB and matter power spectra, 
as these would be altered if there were a higher gravitational wave energy density at the
time of decoupling. The CMB upper limit \cite{PhysRevLett.97.021301} is
\begin{equation}
\int \Omega_{\rm GW}(f) d\left( \ln{f}\right) < 1.3\times 10^{-5}.
\end{equation}
Our upper limit is not sensitive enough to improve on these indirect upper 
limits, however, these indirect bounds only apply to a background of 
cosmological origin, whereas the bound presented here applies to astrophysical signals as well.

\section{Conclusions}
\label{sec:conclusion}

Data acquired by the LIGO and Virgo interferometers 
have been analysed to search for a stochastic background of
gravitational waves. 
This is the first time that data from LIGO and Virgo have been used jointly
for such a search, and we have demonstrated that the addition of 
Virgo increases the sensitivity of the search significantly, reducing the
error bar by $23\%$ even though
the length of time for which Virgo was taking data was approximately 
one fifth of the time of the LIGO run. 
The upper limit obtained with the LIGO interferometers only 
is the most sensitive direct result in this frequency band to date, 
improving 
on the previous best limit, set with the joint analysis of ALLEGRO and LIGO 
data, by a factor of $\approx 7$. 

Adding Virgo improves the sensitivity 
across the frequency band, largely due to the 
the addition of pairs which have different overlap reduction 
functions. This enables us to cover the frequency band more evenly, as well
as effectively increasing the total observation time.
We can
see that the sensitivity of the search is much improved by adding Virgo by
comparing the standard deviations in Table \ref{t:results}. 
However, in this case, the increased sensitivity did not lead to a decreased
upper limit, as the joint estimator of $\Omega_3$ obtained by the the full LIGO-Virgo search
was higher than the the estimator obtained by the LIGO-only analysis. 

As part of this analysis, we have also developed a method of marginalizing
over the error on the amplitude calibration of several 
interferometers. The methods used in this paper will be useful for future
analyses of data from the network of interferometers, which we expect to grow,
eventually including not only interferometers in North America and Europe, but also 
hopefully around the world. 

The authors gratefully acknowledge the support of the United States
National Science Foundation for the construction and operation of the
LIGO Laboratory, the Science and Technology Facilities Council of the
United Kingdom, the Max-Planck-Society, and the State of
Niedersachsen/Germany for support of the construction and operation of
the GEO600 detector, and the Italian Istituto Nazionale di Fisica
Nucleare and the French Centre National de la Recherche Scientifique
for the construction and operation of the Virgo detector. The authors
also gratefully acknowledge the support of the research by these
agencies and by the Australian Research Council, 
the International Science Linkages program of the Commonwealth of Australia,
the Council of Scientific and Industrial Research of India, 
the Istituto Nazionale di Fisica Nucleare of Italy, 
the Spanish Ministerio de Educaci\'on y Ciencia, 
the Conselleria d'Economia Hisenda i Innovaci\'o of the
Govern de les Illes Balears, the Foundation for Fundamental Research
on Matter supported by the Netherlands Organisation for Scientific Research, 
the Polish Ministry of Science and Higher Education, the FOCUS
Programme of Foundation for Polish Science,
the Royal Society, the Scottish Funding Council, the
Scottish Universities Physics Alliance, The National Aeronautics and
Space Administration, the Carnegie Trust, the Leverhulme Trust, the
David and Lucile Packard Foundation, the Research Corporation, and
the Alfred P. Sloan Foundation. This is LIGO document LIGO-P1000128.

\bibliography{S5ligovirgo}

\end{document}